\documentclass[prl,aps,floats,superscriptaddress,floatfix,twocolumn]{revtex4}
\usepackage{amssymb,amsmath}
\usepackage{amsmath,amssymb}
\usepackage{graphicx}
\usepackage{psfrag}

\def\beq{\begin{equation}}
\def\eeq{\end{equation}}
\def\bea{\begin{eqnarray}}
\def\eea{\end{eqnarray}}
\begin{document}
 \title{Tethered membranes do not remain flat for strong structural asymmetry}
\author{Tirthankar Banerjee}\email{tirthankar.banerjee@saha.ac.in}
\affiliation{Condensed Matter Physics Division, Saha Institute of
Nuclear Physics, Calcutta 700064, India}
\author{Niladri Sarkar}\email{niladri2002in@gmail.com}
\affiliation{Max-Planck Institut f\"ur Physik Komplexer Systeme, N\"othnitzer 
Str. 38,
01187 Dresden, Germany}
\author{Abhik Basu}\email{abhik.basu@saha.ac.in,abhik.123@gmail.com}
\affiliation{Condensed Matter Physics Division, Saha Institute of
Nuclear Physics, Calcutta 700064, India}

\date{\today}
\begin{abstract}
 We set up the statistical mechanics for a nearly 
flat, thermally equilibrated fluid membrane, attached to
 an elastic network through one of its sides. We {
predict that} the resulting structural
 (inversion) asymmetry 
of the membrane, notably due to the elastic network attached to one of 
its sides, can generate a local spontaneous curvature $C_0$, that may in turn
 destabilize the otherwise flat membrane. As $C_0$ rises above a threshold 
at a fixed temperature, a flat tethered 
 membrane  in the 
thermodynamic limit becomes structurally unstable, 
 {signaling {\em crumpling}  of the flat membrane}.
 In-vitro experiments on 
  red blood cell membranes after depletion of adenosine-tri-phosphate molecules 
and artificial deposition of spectrin filaments on 
lipid bilayers  may be used to verify our results.
\end{abstract}

\maketitle

Statistical flatness, a well-known feature of inversion-symmetric tethered or 
polymerized membranes at 
sufficiently low 
temperature ($T$) is marked by 
orientational long range order (LRO)~\cite{weinberg}.
 Examples of tethered membranes are plentiful 
covering
biological~\cite{Pontes, Giess, munro}, 
physical~\cite{wiese,md-tethered,mcs,moldovan}
 to chemical~\cite{sinner} systems. Red blood cell (RBC) membranes 
are 
one of the most well-known biological realizations of polymerized 
membranes~\cite{joanny1, joanny2}.
 Statistical properties of inversion-symmetric tethered membranes have been
extensively studied by now, see, e.g., Refs.~\cite{weinberg,chaikin}. { For 
instance, these membranes show a low-$T$ statistically flat 
phase~\cite{weinberg,chaikin}, and a second order crumpling transition to a 
high-$T$ crumpled phase~\cite{weinberg,peliti2}. Furthermore, the scaling 
exponents that characterize the small fluctuations in the low-$T$ flat phase 
have been calculated within perturbative renormalization group (RG) 
framework.} These 
are however idealizations of more general
inversion-asymmetric tethered membranes.  For instance, both {\em 
in-vivo} RBC membranes and {\em in-vitro} spectrin-deposited model 
lipid bilayers are {\em structurally} or {\em inversion asymmetric}, owing 
to the attachment of the elastic network on one side of the membrane. These 
have no theory to date.
Understanding how structural asymmetry affects the statistical properties of 
nearly flat tethered 
membranes forms the principal motivation of this study.

\begin{figure}[htb]
\includegraphics[width=8cm]{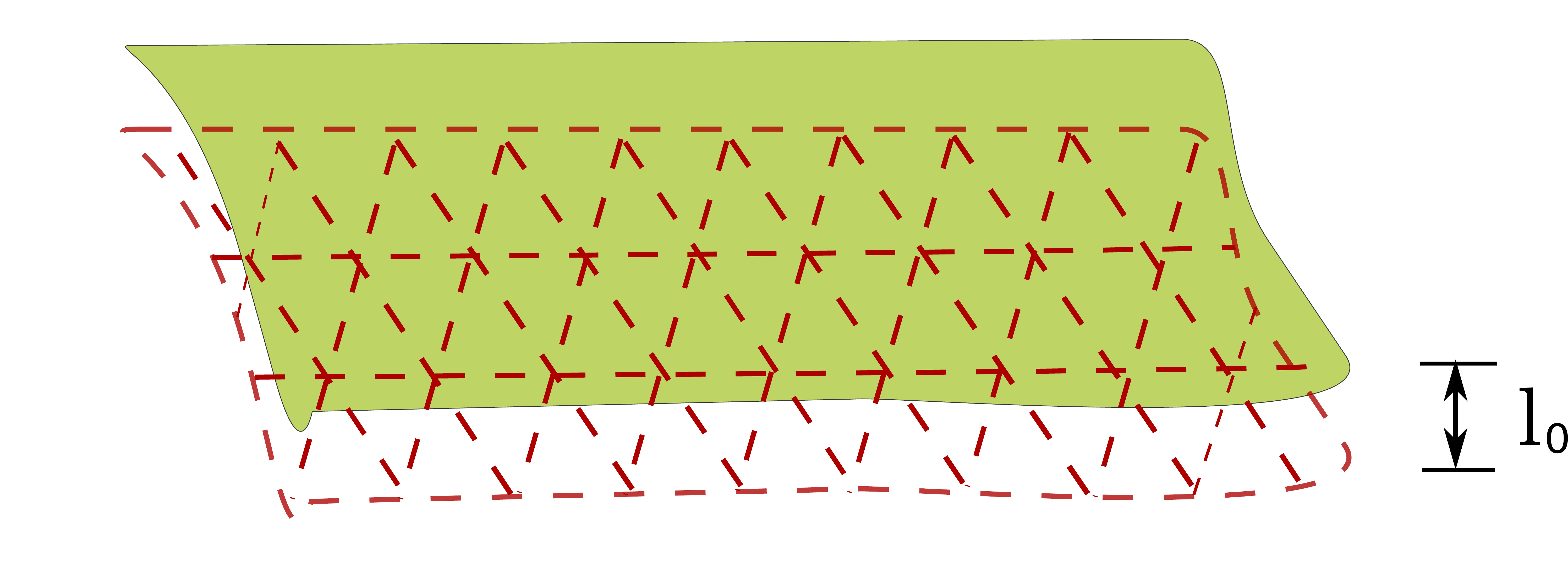}
\caption{(color online) Schematic top view of a membrane (black quadrangle) 
coupled to an
elastic network (broken red triangular lattice) on the bottom side, $l_0$ is 
the average distance between the two.  The membrane and the 
elastic network are joined at the lattice points (see, e.g., 
Ref.~\cite{alberts} for RBC membrane structures, not shown in this diagram); see 
text.  }  \label{model}
\end{figure}

 Here we construct a coarse-grained continuum model for a fluid 
membrane 
attached to an elastic network
 on one side, in thermal equilibrium; see Fig.~\ref{model} for schematic model 
diagrams. We use it to investigate the effects of 
asymmetry on membrane conformation 
fluctuations. We uncover  a novel structural instability in the 
flat membrane at fixed 
$T$, controlled by the local 
strain-dependent spontaneous 
curvature $C_0$, a direct measure of the degree of asymmetry. {This 
indicates a novel asymmetry-induced crumpling of the 
membrane~\cite{weinberg,luca}.}
 Our results are generic in nature and can be tested in adenosine-tri-phosphate 
(ATP) depleted RBC membranes in equilibrium or
 {\em in-vitro} deposition of spectrin filaments on a model lipid 
bilayer~\cite{artificial}. {In addition, our theory should be a starting point 
to 
construct a generic hydrodynamic description for  
live RBC membranes~\cite{joanny1,joanny2}.}

 In order to construct a minimal coarse-grained model designed to extract 
the essential 
physics of the problem,
 we consider a tensionless fluid membrane with a bending modulus $\kappa_0$ 
in thermal equilibrium. The fixed connectivity spectrin network, attached to 
one side 
of the 
membrane, is treated as an elastic continuum parametrized 
by the appropriate Lam\'e constants or elastic modulii $\mu,
 \lambda>0$~\cite{weinberg,chaikin}  in the long wavelength limit (valid over 
length
scales $\gg$ typical spectrin mesh size~$\sim 50 nm$~\cite{safran}) .
 In stark contrast to their symmetric counterparts, we show that nearly flat
 asymmetric tethered membranes in equilibrium { becomes} {\em structurally 
unstable 
} yielding a  {\em crumpled state}, controlled by $C_0$. 
 From our theory, we show that 
  the membrane  
 (i) always remains statistically flat with LRO in thermodynamic limit (TL) 
for low $C_0$, 
(ii) stays flat for system size $L$ smaller than a persistence length $\xi$ 
(see below)
 and {becomes  unstable} for $L>\xi$, for an intermediate 
range of $C_0$, { implying 
a diverging $\langle C_0\rangle$ for $L>\xi$, and (iii) 
gets {unstable}
 for any $L$, large or small, for large enough $C_0$.}

{ We consider the spectrin layer - lipid membrane interaction in a 
{\em strong coupling} limit, i.e., strong interactions without any dissociation 
between them~\cite{joanny}.}
 General symmetry considerations (i.e., invariance under translation and 
rotation) then
 dictate the form of the free energy functional $\mathcal F$ for a nearly flat 
asymmetric tethered membrane. 
 In the coarse-grained long wavelength
 limit, we describe the membrane conformations by a single-valued field $h({\bf 
r})$  in the Monge 
gauge and lateral displacement by a two-dimensional ($2d$) vector field ${\bf 
u}({\bf 
r})$~\cite{weinberg,chaikin}. {For simplicity we assume a fixed 
distance $l_0$ between the lipid membrane and the spectrin network; we 
ignore 
self-avoidance, and any relative motion between the spectrin network and the 
lipid membrane. We also ignore any defect, e.g., missing bonds in the spectrin network.} 
Then, ${\mathcal{F}}$ takes the form
 \begin{eqnarray}\label{free energy}
 {\mathcal{F}} &=& \frac{1}{2}\int d^2 r [\kappa_0(\nabla^2 h)^2+ 
\lambda u_{ii}^2 
\nonumber \\ &+& 
2\mu u_{ij}u_{ij} + 2\chi u_{ii}\nabla^2 h],
\end{eqnarray}
to the leading order in gradients; $r=|{\bf 
r}|$, ${\bf r}=(x,y)$ with (${\bf 
r}, h$) denoting the coordinate of a point on the membrane in the 
three-dimensional embedding space.
 Here  
 strain tensor $u_{ij}=\frac{1}{2}(\nabla_i u_j + \nabla_j u_i +\nabla_i h 
\nabla_j h)$, ignoring terms quadratic in $\nabla_i u_j$, {which 
are irrelevant here in a scaling sense}~\cite{chaikin}. 
In (\ref{free 
energy}) we have included a  generic 
inversion-symmetry breaking term $\chi \nabla^2 h u_{ii}$, 
that couples local compressibility of the network with the local mean 
curvature. In the symmetric 
limit 
$\chi=0$; see Ref.~\cite{safran} for a model of RBC in terms of a solid 
and fluid membrane without the $\chi$-term.
Free energy 
$\mathcal F$
 implies a 	local spontaneous curvature $C_0=(\chi/\kappa_0) u_{ii}$, which 
scales with $\chi$ and naturally vanishes 
 in the symmetric limit. A larger $\chi^2$ signifies that 
a local spectrin compressibility induces a larger local mean curvature.
Parameter $\chi$ can be positive or negative; a 
reversal in the sign of $\chi$ merely reverses $C_0$. 
We show below that $\chi^2$, 
or equivalently, the magnitude of $C_0$ controls the crumpling of an otherwise 
flat membrane.

 We resolve $\bf u$ as $u_i=u_i^{L}+u_i^{T}$, where $u_i^{L}=q_iq_ju_j/q^2$ 
and $u_i^{T}
 =P_{ij}u_j$ are longitudinal
 and transverse components of $u_i$, respectively, for wavevector
$\bf q$; 
 $P_{ij}({\bf q})=\delta_{ij}-q_iq_j/q^2$ is the transverse projection 
operator~\cite{operator}, $i,j=x,y$.
  Thus up to the bilinear order in fields, free energy (\ref{free energy}) takes 
the form 
 \begin{eqnarray}
  {\mathcal F}_g &=& \int \frac{d^2 q}{(2\pi)^2} [\frac{\kappa_0}{2}q^4 |h({\bf 
q})|^2 
  + (\frac{\lambda}{2}+\mu)\{q^2|u^{L}({\bf q})|^2\nonumber \\ &-& 
\frac{2iq\chi 
h({\bf q}) u^{L}({\bf -q})}{2\mu+\lambda}\}+\mu 
q^2|u^{T}({\bf q})|^2],\label{bilinfree}
 \end{eqnarray}
 where, $h({\bf q})$, $u^L({\bf q})$ and $u^T({\bf q})$ 
 are the Fourier transforms of $h({\bf x})$
 and the magnitudes of $u_i^L({\bf x})$ and $u_i^T({\bf x})$, 
respectively. 
 Fields $u^T({\bf q})$ and $u^L({\bf q})$ in the partition function ${\mathcal 
Z}=\int {\mathcal D}h{\mathcal D} u^T{\mathcal D}u^L\exp(- F)$ (with $K_BT=1$, 
$K_B$ is the Boltzmann's constant) may be integrated out exactly 
 to obtain an effective 
 free energy functional that depends only on $h({\bf q})$, and thence an 
effective bending modulus
 $\kappa$:
 \begin{equation}\label{kappa}
  \kappa=\kappa_0-\frac{\chi^2}{2\mu+\lambda},
 \end{equation}
  see Appendix (AP) for details. 
  Evidently, $\kappa<\kappa_0$. 
Thermodynamic stability 
of an assumed flat tensionless 
  membrane clearly requires $\kappa>0$, else instability ensues. 
  Equation~(\ref{kappa}) then yields a threshold for $\chi$ given by 
  $\chi_U^2=\kappa_0(2\mu+\lambda)$, 
  above which $\kappa<0$ for all $q$ and,  {a flat membrane 
becomes {\em crumpled} independent of its size~\cite{weinberg,luca}.}


 How nonlinear effects may modify the above results remains to be 
seen. 
Since ${\mathcal F}$
 is bilinear in $u_{i}$, we can integrate over $u_{i}$ in (\ref{free 
energy}) {\em exactly} to arrive at an effective free energy $F_h$
 that depends only on $h$( now including the nonlinear contribution to 
$u_{ij}$):
  \begin{eqnarray}\label{free energy_h}
  {\mathcal F}_h &=& \frac{1}{2}\int d^2 r [\kappa(\nabla^2 h)^2 +
  \frac{A}{4}(P_{ij}\nabla_i h \nabla_j h)^2 \nonumber \\
  &+& B (\nabla^2 h) (P_{ij} \nabla_i h \nabla_j h)],
 \end{eqnarray}
 where, $A=\frac{4\mu(\mu+\lambda)}{2\mu+\lambda}$
and $B=\frac{2\chi\mu}{2\mu+\lambda}$ are coupling constants
 in the effective theory. Coupling $A$ is positive by construction and is 
responsible for the low-$T$ flat phase, while  $B$, 
being linear in $\chi$, can be both positive and negative. 
 For a full derivation of ${\mathcal F}_h$, see AP. Notice that $B$ 
changes sign for $h \rightarrow -h$, and thus encodes the asymmetry
 in the nonlinear theory; $B=0$ in the symmetric limit for which 
 our model reduces to that of a symmetric tethered membrane in 
equilibrium~\cite{chaikin}. 
The $B$-term in (\ref{free energy_h}) may be interpreted as 
interacting mean 
and Gaussian curvatures via long range interactions; see AP. This is analogous 
to the 
interpretation of the nonlinear 
term with coefficient $A$, as 
long range interactions between local 
Gaussian curvatures in the membrane~\cite{weinberg}.

 \par
 {Nonlinear $A$- and $B$-terms in ${\mathcal F}_h$ necessitate perturbative 
approaches to the present study.}
At the 
one-loop order (equivalently, to the lowest orders in $A$ and $B$), $\kappa_0$ 
receives two fluctuation corrections, each 
originating from non-zero $A$ and $B$, respectively; see AP for details. We 
find for the $q$-dependent renormalized bending modulus $\kappa_R (q)$, 
and $\delta\kappa =\kappa_R(q)-\kappa$,
\begin{eqnarray}
 \delta\kappa&=&\frac{A}{\kappa}\int \frac{d^2q}{(2\pi)^2} 
\frac{[\hat q_i P_{ij}({\bf q}_1)\hat q_j]^2}{|{\bf q+q}_1|^4} - 
\frac{B^2}{2\kappa^2}\int \frac{d^2q}{(2\pi)^2}\hat q_i 
P_{ij}({\bf q}_1)\hat q_j\nonumber\\&&\times [\frac{\hat q_i P_{ij}({\bf 
q}_1)\hat q_j}{|{\bf q+q}_1|^4} + \frac{ \hat q_m P_{mn}({\bf q+q}_1)\hat 
q_n}{q_1^4}], \label{mct}
\end{eqnarray}
$\hat {\bf q}$ is the unit vector along $\bf q$.
Both the integrals on the rhs of (\ref{mct}) diverge as $1/q^2$ for small $q$. 
The 
former, existing for both symmetric~\cite{weinberg,chaikin} and asymmetric 
tethered membranes contributes positively to $\kappa$, where as the latter, 
that exists only in asymmetric membranes, contributes negatively. 
Evidently, the stretching energy drastically enhances $\kappa_R(q)$ for small 
$q$, for positive rhs of (\ref{mct})~\cite{nelson-peliti}; for 
negative rhs, the effect is just the opposite. Assuming net positive 
corrections to $\kappa$ in (\ref{mct}), a simple 
self-consistent theory  unsurprisingly yields 
$\kappa_R(q)\sim 1/q$~\cite{nelson-peliti}. More sophisticated approaches that 
systematically handles the diverging corrections as in (\ref{mct}) and accounts 
for the fluctuation corrections (if any) to the nonlinear couplings $A$ and $B$
are based on the framework of perturbative Wilson momentum shell 
renormalization group (RG)
 technique~\cite{chaikin}, together with an $\epsilon$-expansion, where 
$\epsilon=4-d$ {with $(d+1)$ referring to the embedding space dimension}  
 (see AP for some technical details). 
 To this end, we eliminate fields $h({\bf q})$ (where 
$\Lambda/b <q<\Lambda,
 b>1$) by integrating perturbatively up to the one-loop order; 
 $\Lambda$ is an upper wave-number cut-off. This is followed by
 a rescaling of wave-vectors ${\bf q}$ via ${\bf q}^{\prime}=b{\bf q}$ 
 and the field $h({\bf 
q})=\zeta_h h^{\prime} ({\bf q}^{\prime})$;
 $\zeta_h=b^{(d+4-\eta)/2}$, $\eta$ being the anomalous dimension of $h({\bf 
q})$ (yet unknown). 

\begin{figure}[htb]  
 \includegraphics[width=10cm]{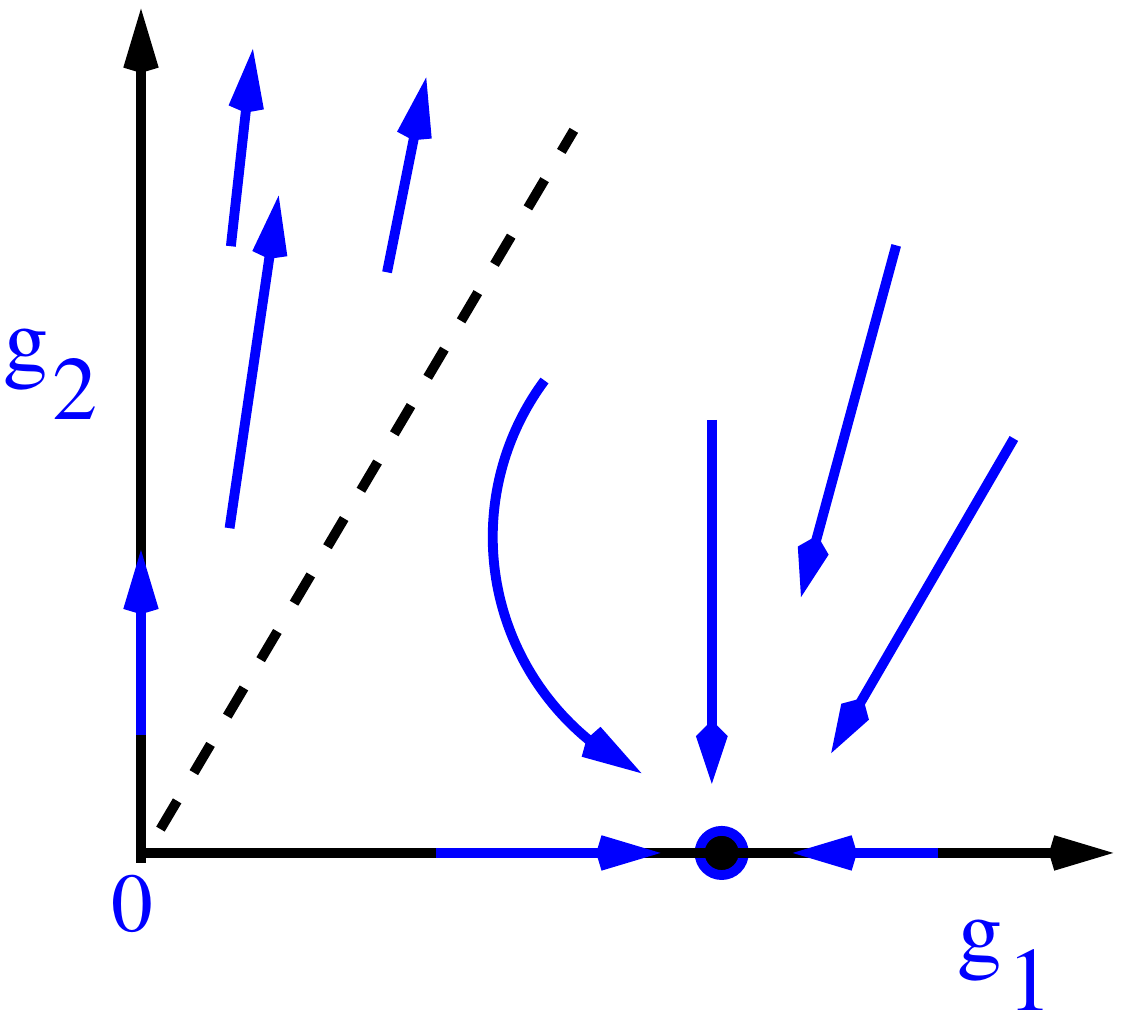}
 \caption{(Color online)Schematic flow lines in the $g_1-g_2$ plane. Small filled circle 
(blue) marks the stable FP ($2\epsilon/5,0$). The broken (nonuniversal) line is 
a schematic 
demarcation between the region controlled by the stable FP 
(corresponding to a flat membrane) and the region where flow 
lines 
point towards infinity, signalling breakdown of a flat membrane.}  
\label{fp}
 \end{figure}

  Assuming again net positive corrections to $\kappa$, with $b=\exp[ 
l]$, the recursion relation~\cite{chaikin} for $\kappa$ takes the form
\begin{equation}
 \frac{d\kappa}{dl}=\kappa[-\eta + K_d(\frac{A}{\kappa^2}
   -\frac{2 B^2}{\kappa^3})],\label{renor_kappa}
\end{equation}
 where $K_d=\frac{d^2-1}{d(d+2)}$. 
 We now define two effective coupling 
constants,
 $g_1=\frac{A}{(2\pi)^d\kappa^2}$ and $g_2=\frac{B^2}{(2\pi)^d \kappa^3}$, with
 \begin{eqnarray}
 \frac{dg_1}{dl}=g_1[\epsilon -\frac{5g_1}{2}+4g_2],
 \frac{dg_2}{dl}=g_2[\epsilon-4g_1+6g_2],\label{flow2}
\end{eqnarray}
as the respective RG flow equations; see AP.
At 
the RG fixed points 
(FP), $dg_1/dl=0=dg_2/dl$ yielding $g_1=2\epsilon/5,\,g_2=0$ as the only 
{ globally} stable FP. This 
yields $\eta=2\epsilon/5$~\cite{mismatch}, {which 
corresponds to LRO}. In other words, 
asymmetry is {\em 
irrelevant} (in a 
scaling/RG sense) { in the flat phase}.
Now consider the stability of the FP
 in the $g_1-g_2$ plane; see Fig.~\ref{fp}. Notice that the unstable FP 
($g_1=0,\,g_2=0$) is globally unstable; i.e., unstable along the 
$g_2$-direction as well. In general flow 
equations (\ref{flow2}) suggest that with initial conditions 
$g_2(l=0)\gg g_1(l=0)$, the flow lines {\em do not} flow to the stable FP; 
instead they appear to flow to infinity, signalling breakdown of a flat 
membrane. This is 
shown schematically in Fig.~\ref{fp}.

\par
\begin{figure}[htb]  
 \includegraphics[width=7.6cm]{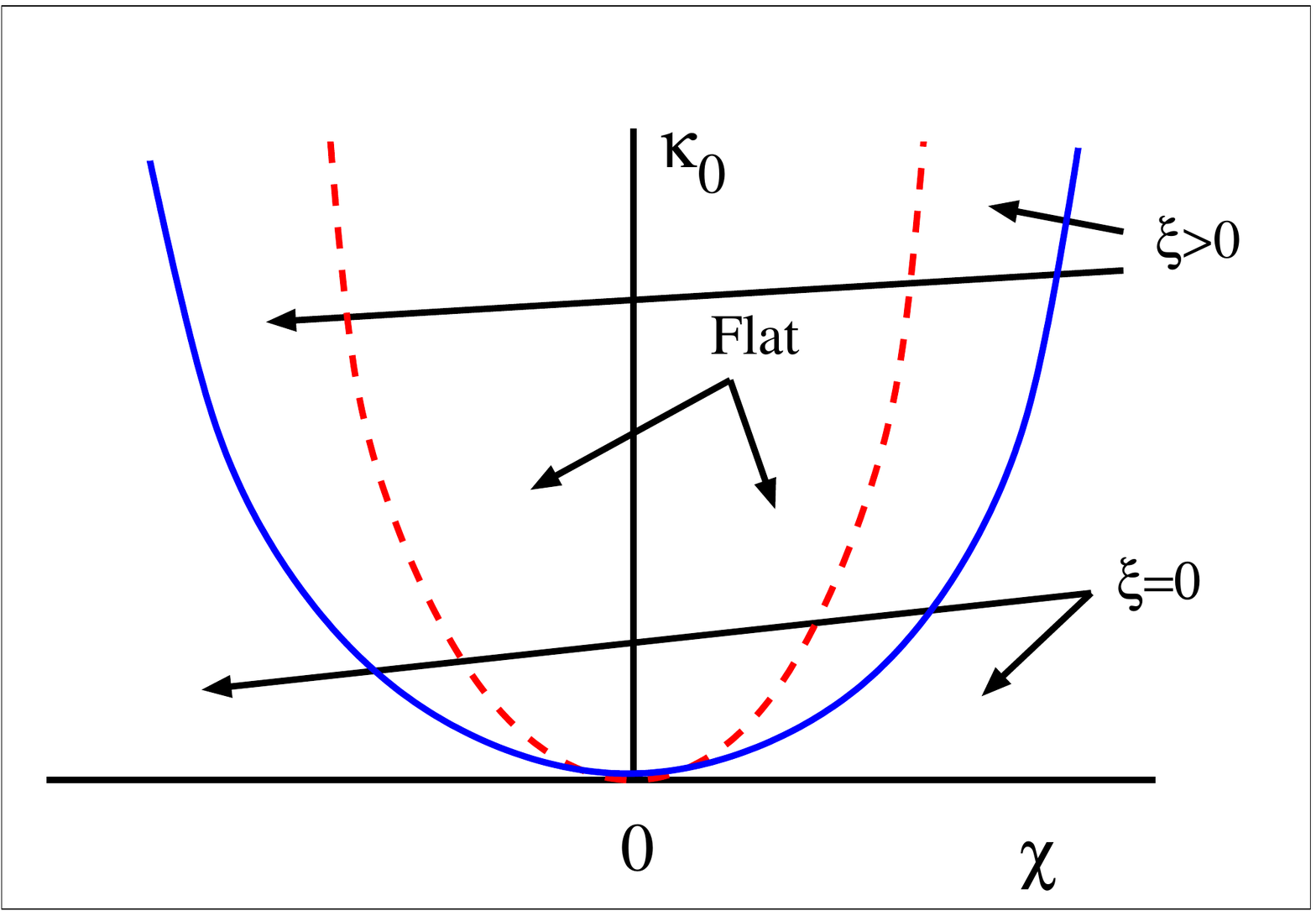}\hfill
 \caption{(Color online) Schematic phase diagram in the $\chi-\kappa_0$ plane. The red, inner 
parabola (broken line) and the blue, outer 
parabola (continuous line) represent, respectively, $\chi^2=\chi_L^2$ and 
$\chi^2=\chi_U^2$. {In the region $\chi_L^2<\chi^2<\chi_U^2$, the membrane
becomes unstable for a length $L>\xi$ (finite), the persistence length. Inside
the inner parabola, $\xi \rightarrow \infty$ and the membrane always remains flat.
Beyond the outer parabola ($\chi^2 > \chi_U^2$),
the membrane no longer remains flat at any $L$ (see text).}}
\label{k-chi}
 \end{figure}

 Let us now consider the 
consequence of negative  $\kappa_e=\kappa_R(q\sim 2\pi/L)<0$ possible 
for 
sufficiently large $B^2$, for a membrane of 
size $L$. Keeping the dominant corrections {we obtain at 2d},
\begin{equation}
 \kappa_e-\kappa \approx \frac{3}{8}(\frac{A}{\kappa}
 -\frac{2 B^2}{\kappa^2}) \int_{2\pi/L}^{\Lambda} \frac{d^2 q_1}{(2\pi)^2 
q_1^4}.\label{crumplekappa}
 \end{equation}
A membrane with a size larger than $\xi$
can no 
longer remain flat and  {\em destabilizes} or {\em crumples} for 
$\kappa_e(L=\xi)=0$, 
 yielding
(neglecting terms $O(1/\Lambda^2)$ which are small for $\xi\gg 1/\Lambda$)
 \begin{eqnarray}\label{L_chiEq}
  &&\frac{3 K_B
T\xi^2}{128\pi^3}\left[A\left(\kappa_0-\frac{\chi^2}{2\mu+\lambda}
\right)-\frac { 8\chi^2\mu^2}{(\lambda+2\mu)^2}\right]
 \nonumber \\&=&-\left(\kappa_0-\frac{\chi^2}{2\mu+\lambda}\right)^3,
 \end{eqnarray}
  { where an explicit factor of $K_B T$ has been inserted in (\ref{L_chiEq}). 
Thus, $\xi\rightarrow\infty$ as $\chi^2\rightarrow \chi_L^2$, a lower 
threshold where}
\begin{equation}\label{chi_c eq}
 \chi_L^2=\frac{A\kappa_0  
(2\mu+\lambda)^2}{A(2\mu+\lambda)+8\mu^2}=\frac{\kappa_0 
(2\mu+\lambda)}{1+\frac{2\mu}{\mu+\lambda}}.
\end{equation}
For $\chi^2 > 
\chi_L^2$, $\kappa_e (\xi)=0$, leading to destabilization  of the flat membrane 
{and its {\em crumpling}}
at scales 
larger than $\xi$. On the other hand, $\xi\rightarrow 0$ as $\chi^2\rightarrow 
\chi_U^2=\kappa_0(2\mu+\lambda)$, such that for $\chi^2>\chi_U^2$, the 
membrane crumples at all scales. Unsurprisingly, $\chi_U^2$ is 
 same as obtained from Eq.~(\ref{kappa}) above; clearly 
$\chi_U^2>\chi_L^2$ {with none of them having any $T$-dependence.} {For 
$\chi^2<\chi_L^2$, a flat membrane ensues in TL.}
 See Fig.~\ref{k-chi} for a schematic phase diagram in the $\chi-\kappa_0$ 
plane.
 Furthermore from Eq.~(\ref{L_chiEq}), 
 $\xi=\sqrt{\frac{128\pi^3\kappa^3}{K_B T(6B^2-3A\kappa)}}$, yielding $\xi\propto 
1/\sqrt T$ over a relevant range of $T$. 

 
Consider now the mean spontaneous curvature $\langle C_0\rangle$ in 
the different regimes  delineated by $\chi$. For 
 $\chi^2<\chi_L^2$, a flat membrane ensues in the long wavelength 
limit, 
and  $\langle C_0\rangle=0$; see AP. This is consistent with a statistically 
flat membrane in TL. {In contrast, for $\chi^2>\chi_L^2$, $\langle C_0\rangle =\langle 
({\boldsymbol\nabla}h)^2\rangle\chi/\kappa_0=\int_{2\pi/L}^{\Lambda} \frac{d^2q}{(2\pi)^2} q^2\langle 
|h({\bf q})|^2\rangle\chi/\kappa_0$ clearly diverges for $L>\xi$}. Thus, {
the crumpling
instability} is 
generically associated with a diverging spontaneous curvature.


A finite surface 
tension $\sigma$, if present, will control the long wavelength fluctuations of 
the membrane~\cite{surface-tension}. 
We note that $\sigma$ receives no relevant fluctuation corrections 
from the nonlinear terms $A$ and $B$ in (\ref{free energy_h}).  Interestingly 
however, for a 
sufficiently large $\chi^2$,  $\kappa_e$ that dominates at the intermediate 
wavevectors, 
may become negative, leading to finite wavevector instabilities, a feature 
testable in controlled {\em in-vitro} experiments.
 
  Although a full RBC~\cite{alberts} is clearly 
symmetric under 
inversion (since the two lipid bilayers are identical in structure), each 
bilayer clearly has an asymmetric environment having the spectrin network only 
on one (inner) side of it. Following the logic outlined above, our theory can 
be readily extended to two identical lipid membranes confining a spectrin 
network, that may serve as a minimal model for a whole RBC. Using symmetry 
arguments as above, one can construct a coarse-grained free energy. An analysis 
similar to the one above indicates towards statistical flatness of RBC membranes 
for sufficiently low 
spectrin-lipid interaction strength, beyond which crumpling of the RBC should 
be observed. Details will be considered elsewhere~\cite{tb-future}. Recent 
studies on live RBC membranes~\cite{joanny1,joanny2} 
reveal enhanced fluctuations in the RBC membranes for low frequencies. While 
 a live RBC membrane is an {\em active} 
or driven system, and hence outside the scope of our theory, any consistent 
hydrodynamic theory for live RBC membranes should reduce to our theory in its 
equilibrium limit.
 {Our theory reveals the crumpling instability induced by strong enough 
asymmetry 
($\chi^2>\chi_U^2$) in a flat membrane. It must be generalized to study 
the nature of the thermodynamically stable structurally crumpled 
state  ($\chi^2>\chi_U^2$). Formal similarities between a direct generalization 
of 
our 
model to study the $T$-driven crumpling 
transition~\cite{weinberg,peliti2} and the standard Landau theory for 
phase transitions with cubic nonlinearities open up the intriguing possibility 
of new universal scaling at the second order crumpling transition and also 
possibly first order crumpling transitions. Nonetheless, how the structural 
crumpling elucidated here is connected to the well-known thermal crumpling of 
flat tethered membranes~\cite{weinberg,luca} still remains  unresolved.
}

  Independent of the precise values of the model parameters (which may yet 
be unknown experimentally), the general form and structure of the phase diagram 
(\ref{k-chi}) 
   can be tested in non living (ATP-depleted) RBC membrane 
extract~\cite{artificial}, or
 for model asymmetric membranes by binding spectrin to lipids by presenting positive charges to lipid
 surfaces~\cite{toole}. The binding of the elastic network to the membrane surface may be controlled by proteins, e.g.,
 Stomatin~\cite{grzybek}; see also ~\cite{hendrich} for other spectrin-lipid 
interactions. 
In laboratory-based controlled {\em in-vitro} experiments, the 
spectrin-lipid membrane interactions, modeled by $\chi$ here, may be controlled 
by adding cholesterol~\cite{abhijit} in artificially prepared samples, or in 
ATP-removed RBC membranes. Numerical simulations of the 
analogous discrete models of asymmetric tethered membranes, similar to the 
studies in Refs.~\cite{md-tethered,mcs,peng}, should be employed to  verify our 
results.


Our assumptions of a fixed distance between the elastic network and the 
attached lipid membrane is clearly an idealization; for an RBC membrane 
$l_0\sim 30\,nm$ is the average 
distance between the two~\cite{safran}. Small fluctuations 
in $l_0$ about its mean, expected in ATP-depleted RBC membranes, is not 
expected to modify our results in any significant 
manner; see AP for details.                                         
Self-avoidance, 
ignored here for a putative flat membrane, may be important in 
the unstable phase~\cite{weinberg}. 
 Recent studies on thin, 
spherical shells indicate that large enough shells can be crushed by thermal 
effects~\cite{nelson-new}. How the compressibility-mean curvature coupling 
introduced here affect this result may be investigated in future.
The geometric 
nonlinearities associated with the 
Monge gauge are irrelevant (in a scaling/RG sense) and hence have been 
neglected. {We have used a planar geometry for simplicity. In other 
relevant geometries, e.g., spherical (significant for an RBC membrane), there 
is a nonzero mean spontaneous curvature even without any fluctuation that 
characterizes the global shape of the membrane. Whether an asymmetric (i.e., 
one-sided) coupling of a spherical membrane with a spectrin network, described 
via an analog of the $\chi$-term in (\ref{free energy}), either {\em increases} or 
{\em decreases} the mean spontaneous curvature may now depend upon whether the 
spectrin layer is attached to the inner or outer surface of the spherical 
membrane. How that plays out in regard to the instabilities elucidated above is 
an important question that should be studied separately.}
We 
expect our work 
to provide new impetus towards detailed experimental studies of asymmetric 
tethered membranes.

 
 {\em Acknowledgement:-} TB and AB gratefully acknowledge partial 
financial 
support from the Alexander von 
Humboldt Stiftung, Germany under the Research Group 
Linkage Programme (2016).


\newpage
\widetext 
 
 \begin{center}
 \LARGE{APPENDIX (AP)}
 \end{center}
 
 \section{Calculation of ${\mathcal F}_h$ for the full non-linear theory }
 We integrate over $u_{ij}$ to obtain an effective free enrgy that depends only 
on $h$. We proceed by breaking
 $u_{ij}$ into $q$-dependent and $q$-independent parts; ${\bf q}$ being a 
wave-vector, $q=|{\bf q}|$.
 \begin{eqnarray}
  u_{ij}(x) &=& u_{ij}^0 + A_{ij}^0
  +\Sigma_{q \neq 0} [\frac{i}{2}(q_j u_i(q) + q_i 
u_j(q))+A_{ij}(q)] \exp (i {\bf q}.{\bf x}),
 \end{eqnarray}
 where $A_{ij}(q)=\frac{1}{2}\int d^2 x \exp(-i {\bf q}.{\bf x}) (\nabla_i h) 
(\nabla_j h)$, ${\bf x}$
 is a $2d$ Cartesian displacement vector.
 We now use the fact that any $2d$ symmetric second rank tensor can be written 
as a sum of transverse and longitudinal
 parts. Let $\phi(q)=\frac{q_i q_j}{q^2}A_{ij}$. We can write
 \begin{eqnarray}
  A_{ij} &=& \frac{q_i q_j}{q^2}\phi + A_{ij} -\frac{q_i q_j q_m 
q_n}{q^4}A_{mn}\nonumber \\
  &=& \frac{q_i q_j}{q^2}\phi + P_{ij} \Phi,
 \end{eqnarray}
 where, $P_{ij}=\delta_{ij}-\frac{q_i q_j}{q^2}$ and $P_{ij} 
\Phi=[\delta_{mi}\delta_{nj}-\frac{q_i q_j q_m q_n}{q^4}]A_{mn}$.
 Thus $P_{ij}P_{ij}\Phi=\Phi=P_{ij}A_{ij}$. We now write $A_{ij}$ as a 
combination of longitudinal
 and transverse parts :
 \begin{equation}
  A_{ij}=\frac{1}{2}(q_i\theta_j+q_j\theta_i)+P_{ij}\Phi,
 \end{equation}
 where the first term in the rhs represents the longitudinal component of 
$A_{ij}$ and $P_{ij}\Phi$
 gives the transverse part. Here $\theta_i=\frac{q_i\phi}{q^2}$. Thus in real 
space, we can write
 $u_{ij}=\frac{1}{2}(\nabla_i \tilde u_j +\nabla_j \tilde u_i)+P_{ij}\Phi$, 
where $\tilde u_i(x)=u_i(x)+\theta_i(x)$.
 Taking only the $u$-dependent part in ${\mathcal F}$, we have

\begin{equation}
 I = \int d^2 x [\frac{\lambda}{2}u_{ii}^2+\mu u_{ij} u_{ij}+\chi u_{ii} 
\nabla^2 h].
\end{equation}
 Now $\int d^2 x u_{ij} u_{ij}=\int d^2 r[\frac{1}{4}(\nabla_i \tilde u_j 
+\nabla_j \tilde u_i)^2 + (P_{ij}\Phi)^2]$. Also,
 $\int d^2 x u_{ii} u_{jj}= \int d^2 x [(\nabla_i \tilde u_i)^2+ 2 \Phi \nabla_i 
\tilde u_i + \Phi^2]$ and 
 $\int d^2 x (P_{ij} \Phi)^2 = \int d^2 x \Phi^2$.
 We now write $\tilde u_i=\tilde u_i^L+\tilde u_i^T$, where $\tilde u_i^L$ and 
$\tilde u_i^T$
 represent the longitudinal and transverse components of $\tilde u_i$, 
respectively. Since $\tilde u_i^L$ is
 fully longitudinal, we can write $\tilde u_i^L=\nabla_i \psi$, where $\psi$ is 
a scalar. 
 This implies
 
 \begin{eqnarray}
  \int d^2 x (\nabla_i \tilde u_j^L)^2 &=& \int d^2 x (\nabla_i \nabla_j \psi) 
(\nabla_i \nabla_j \psi) = \int (\nabla_i^2 \psi) (\nabla_j^2 \psi)\nonumber \\
    &=& \int d^2 x (\nabla_i \tilde u_i^L)^2
 \end{eqnarray}

 Using these values, we have
 \begin{eqnarray}\label{I-u}
  I = \int d^2 x [\frac{\lambda}{2}[(\nabla_i \tilde u_i)^2+ 2 \Phi \nabla_i 
\tilde u_i + \Phi^2]
     + \mu[(\nabla_i\tilde u_i^L)^2+(\nabla_i \tilde u_j^T)^2+\Phi^2]
+ \chi(\nabla_i \tilde u_i^L+\Phi)\nabla^2 h]
 \end{eqnarray}
 
 We note that the coupling between $u_{ij}$ and $h$ appears only in the form of 
$\tilde u_i^L$ and thus
 $u_i^T$ may be integrated out trivially. After proper recombination of terms in 
Eq.~(\ref{I-u}) of AP, we get
 
 \begin{eqnarray}
  I &=& \int d^2 x [(\lambda/2 + \mu)[(\nabla_i \tilde u_i^L) + \frac{\lambda 
\Phi}{\lambda +2 \mu}+\frac{\chi \nabla^2 h}{\lambda+2\mu}]^2
    - \frac{\lambda^2 \Phi^2}{2\lambda+4\mu}\nonumber \\ &-& \frac{\chi^2 (\nabla^2 
h)^2}{2\lambda+4\mu} - \frac{\lambda \chi \Phi (\nabla^2 
h)}{\lambda+2\mu}
    + (\lambda/2 +\mu) \Phi^2 + \chi \Phi \nabla^2 h]
 \end{eqnarray}

 Integrating over $\tilde u_i^L$, we arrive at the following equation
 
 \begin{equation}\label{I-h}
  I = \int d^2 x [\frac{\mu (\mu+\lambda)}{\lambda/2 + \mu}\Phi^2 + \chi 
(\nabla^2 h) \Phi \frac{\mu}{\lambda/2+\mu}-\frac{\chi^2 (\nabla^2h)^2}{2(2\mu+\lambda)}]
 \end{equation}

 Using Eq.~(\ref{I-h}) above and the value of $\Phi$, we arrive at 
the free energy ${\mathcal F}_h$ of the main text.

\section{Interacting Gaussian and mean curvatures}
 
 Noting that {(see, e.g., Ref.~[1] of the main text)}
\begin{equation}
 -\nabla^2 [\frac{1}{2}P_{ij}(\nabla_i h)(\nabla_j 
h)]=\det\left(\nabla_i \nabla_j h\right),
\end{equation}
yields $P_{ij}(\nabla_ih)(\nabla_jh)=\int d^dx^\prime M(|{\bf 
x}-{\bf x}^\prime|)S({\bf x}^\prime)$, where $S({\bf x})$ is the local Gaussian 
curvature at $\bf x$, $M(|{\bf x}|)$ is the inverse Fourier transform of 
$1/q^2$. Then, 
\begin{equation}
\int d^dx \nabla^2hP_{ij}(\nabla_i h)(\nabla_j h)=\int d^dx 
d^dx^\prime \nabla^2h({\bf x})M(|{\bf x}-{\bf x}^\prime)S({\bf x}^\prime).
\end{equation}
In particular, at $2d$, $M(|{\bf x}|)\sim \ln x$, establishing the picture that 
the $B$-term in the free energy ${\mathcal F}_h$ of the main text may be 
interpreted as interacting mean 
and Gaussian curvatures via long range interactions.

\newpage
\section{One-loop Feynman diagrams}\label{diag}

\begin{figure}
 \includegraphics[width=8cm]{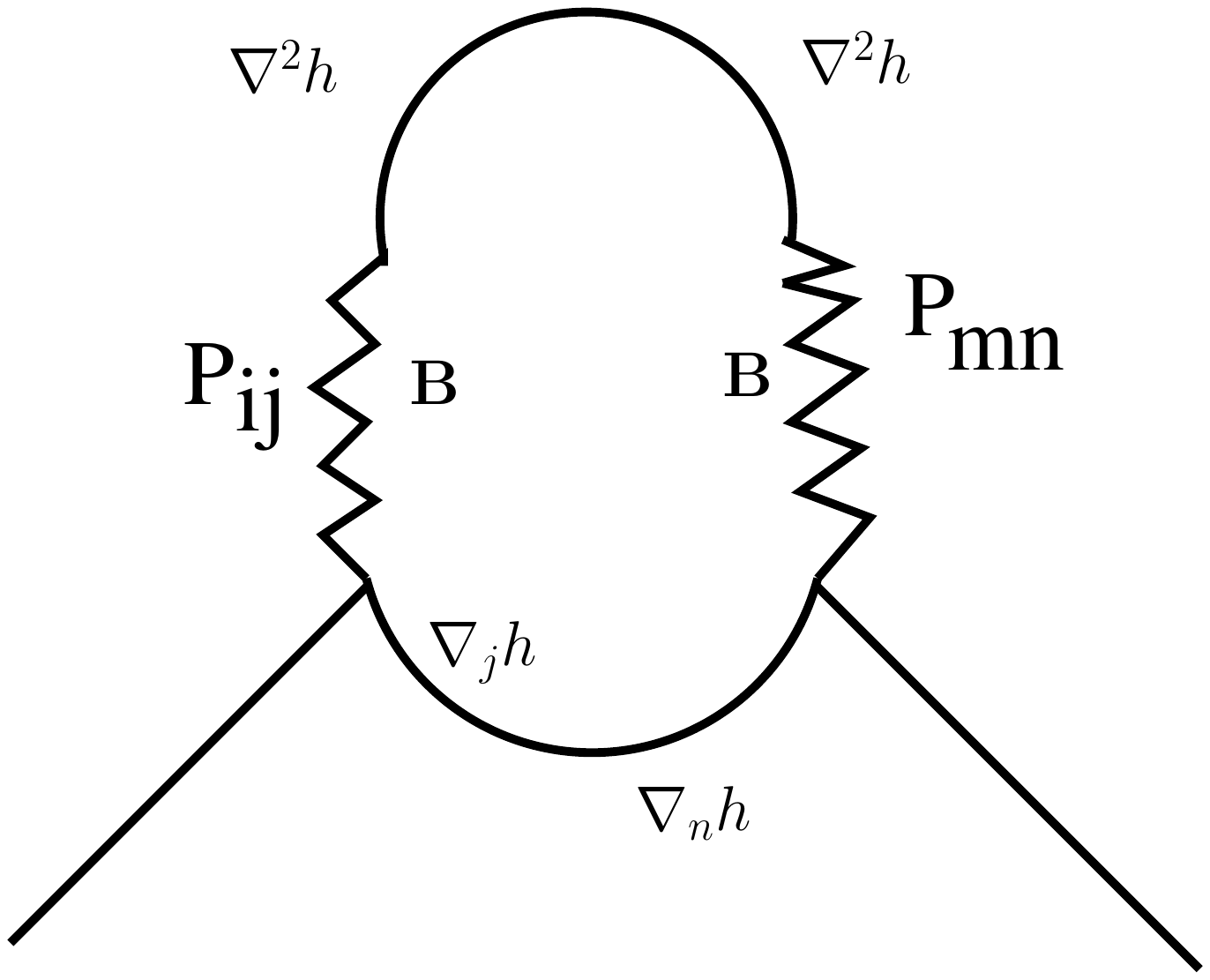}\\
 \includegraphics[width=8cm]{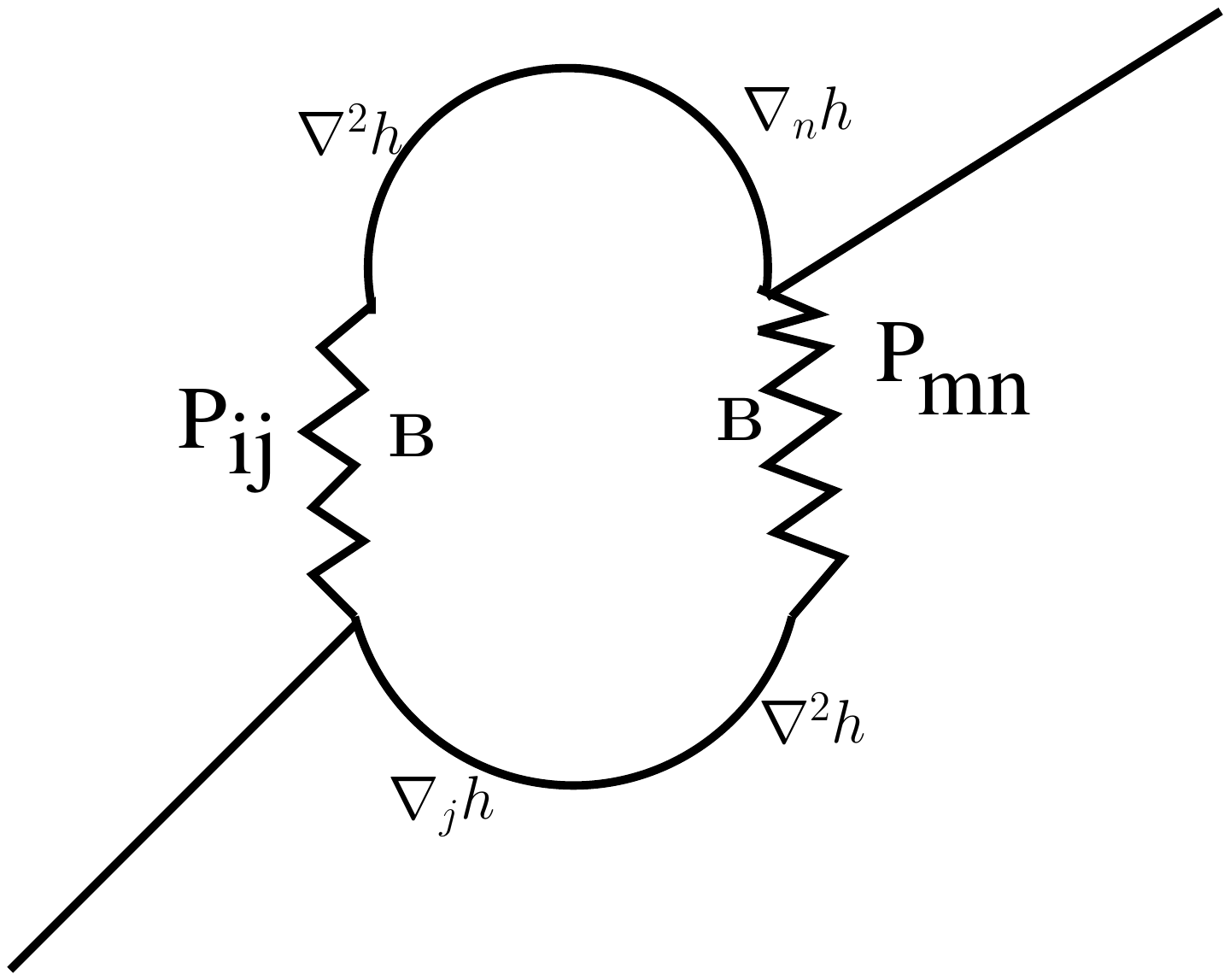}
 \caption{One-loop Feynman diagrams that originate from the 
$B$-nonlinear term and contribute to the fluctuation 
corrections of $\kappa$.
}\label{kappa_dia-all}
\end{figure}

\begin{figure}[htb]
\includegraphics[width=18cm]{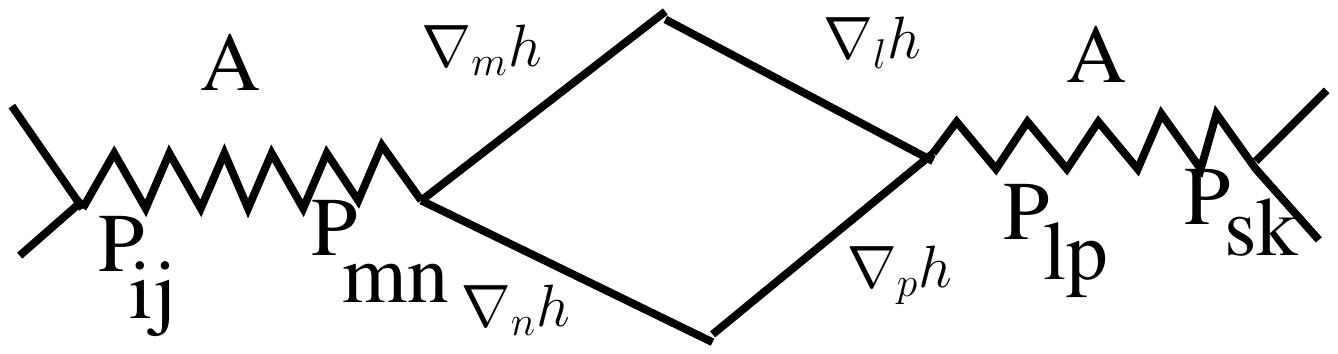}
\caption{One-loop Feynman diagram contributing to fluctuation corrections of 
$A$.}\label{Adiag}
\end{figure}

\begin{figure}[htb]
\includegraphics[width=12cm]{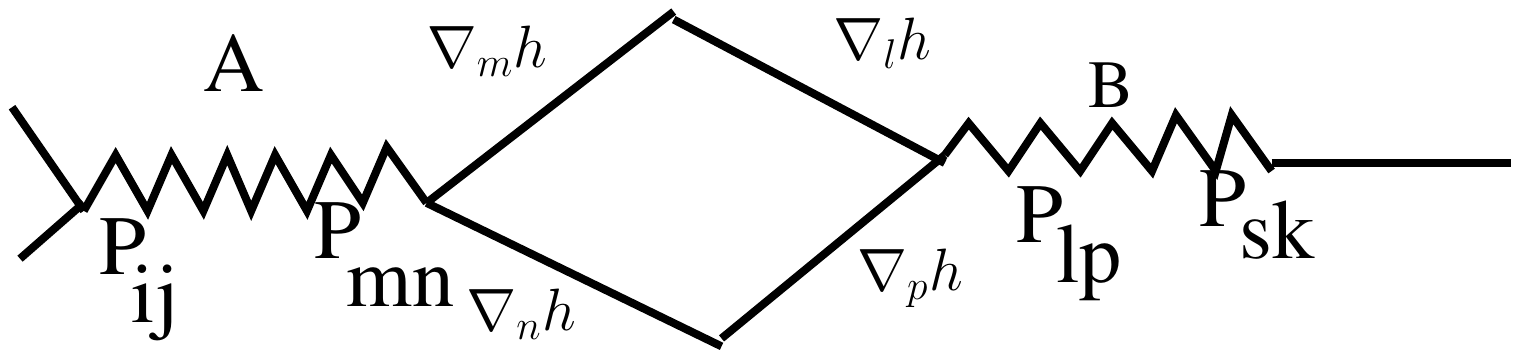}
\caption{One-loop Feynman diagram contributing to fluctuation corrections of 
$B$.}\label{Bdiag}
\end{figure}

The one-loop Feynman diagrams which contribute to the fluctuation corrections 
of $\kappa,A$ and $B$ are shown in Fig.~\ref{kappa_dia-all}, Fig.~\ref{Adiag} 
and Fig.~\ref{Bdiag} of AP, respectively.

\begin{figure}[htb]
 \includegraphics[width=12cm]{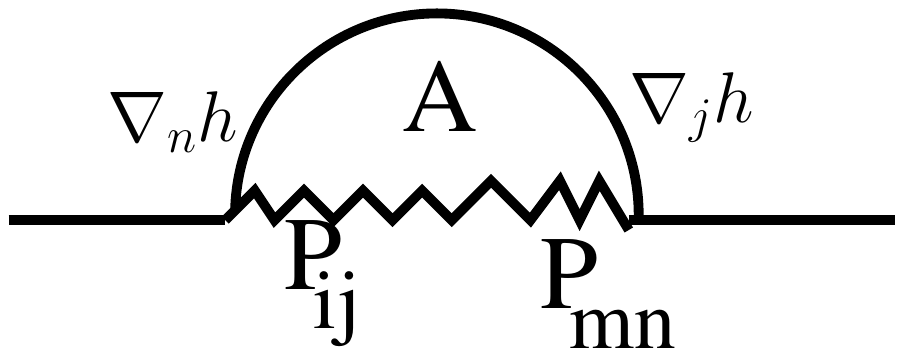}
 \caption{One-loop Feynman diagram that originates from the 
$A$-nonlinear term and contributes to the fluctuation 
corrections of $\kappa$.}
\end{figure}

\newpage

\newpage

\newpage
 \section{Discrete recursion relations}
 
 After rescaling of wave-vectors ${\bf q}$ via ${\bf q}^{\prime}=b{\bf q}$ 
 and the field $h({\bf 
q})=\zeta_h h^{\prime} ({\bf q}^{\prime})$;
 $\zeta_h=b^{(d+4-\eta)/2}$, $\eta$ being the anomalous dimension of $h({\bf 
q})$ 
 
 \begin{eqnarray}
 \kappa^{\prime}&=&b^{-\eta}\kappa[1+ (\frac{A K_d}{\kappa^2}-
 \frac{2 B^2 K_d}{\kappa^3})\int_{\Lambda/b}^{\Lambda} \frac{d^d q}{(2\pi)^d 
q^4}]\nonumber \\
  A^{\prime}&=& b^{-2\eta+4-d}A[1-\frac{A K_d}{2\kappa^2} 
\int_{\Lambda/b}^{\Lambda} \frac{d^d q}{(2\pi)^d q^4}]\nonumber \\
 B^{\prime}&=& b^{\frac{-3\eta+4-d}{2}} B[1-\frac{A K_d}{2 \kappa^2} 
\int_{\Lambda/b}^{\Lambda} \frac{d^d q}{(2\pi)^d q^4}].
 \end{eqnarray}

 The discrete recursion relations in terms of the coupling constants $g_1$ and $g_2$ are given by :
 
 \begin{eqnarray}\label{g_eqns}
  g_1^{\prime}&=& \frac{A^{\prime} K_d}{(2\pi)^d \kappa^{\prime 
2}}=b^{\epsilon} 
g_1 [1- (\frac{5 g_1}{2}
  - 4 g_2) \int_{\Lambda/b}^{\Lambda} \frac{d^d q}{q^4}]\nonumber \\
  g_2^{\prime}&=& \frac{B^{\prime 2} K_d}{(2\pi)^d \kappa^{\prime 
3}}=b^{\epsilon} g_2 [1-(4 g_1
  - 6 g_2) \int_{\Lambda/b}^{\Lambda} \frac{d^d q}{q^4}]
 \end{eqnarray}
 
\section{ Spontaneous curvature and odd order correlators of $h$}

Upon rescaling $h$ by $\sqrt\kappa$, free energy ${\mathcal F}_h$ of the main 
text can be written as
\begin{equation}
 {\mathcal F}_h=\int d^d r [(\nabla^2 h)^2 + (2\pi)^d \frac{g_1}{4} 
(P_{ij}\nabla_i h \nabla_j h)^2 + (2\pi)^dg_2\nabla^2 h (P_{ij}\nabla_i 
h\nabla_j h)],\label{fh}
\end{equation}
in $d$-dimensions. The $g_2$-term in (\ref{fh}) violates the inversion 
symmetry. At the stable RG FP, $g_2=0$, rendering (\ref{fh}) symmetric under 
inversion of $h$. {Mean spontaneous curvature $\langle C_0\rangle$ clearly 
scales with $g_2$ in the effective, long wavelength renormalized theory, and 
hence vanishes in the flat phase in the long wavelength limit.}
Thus, in the renormalized theory all odd order 
correlators of the form $\langle h({\bf 
q}_1) h({\bf q}_2) h({\bf q}_3)....h({\bf q}_n)\rangle,\,{\bf q}_1+{\bf 
q}_2+{\bf q}_3+...+{\bf q}_n=0$ ($n>0$ an odd integer) vanish.  Hence, the 
equilibrium states of an 
asymmetric nearly flat membrane that is {\em thermodynamically stable} is in 
fact {\em 
identical} to their symmetric 
counterparts in  asymptotic long wavelength limit. The equilibrium states of an 
asymmetric tethered membrane in the unstable phase for $\chi^2>\chi_L^2$ are of 
course very 
different from their symmetric 
counterparts.



\section{Effect of fluctuating membrane-network distance}

\begin{figure}[htb]
 \includegraphics[width=14cm]{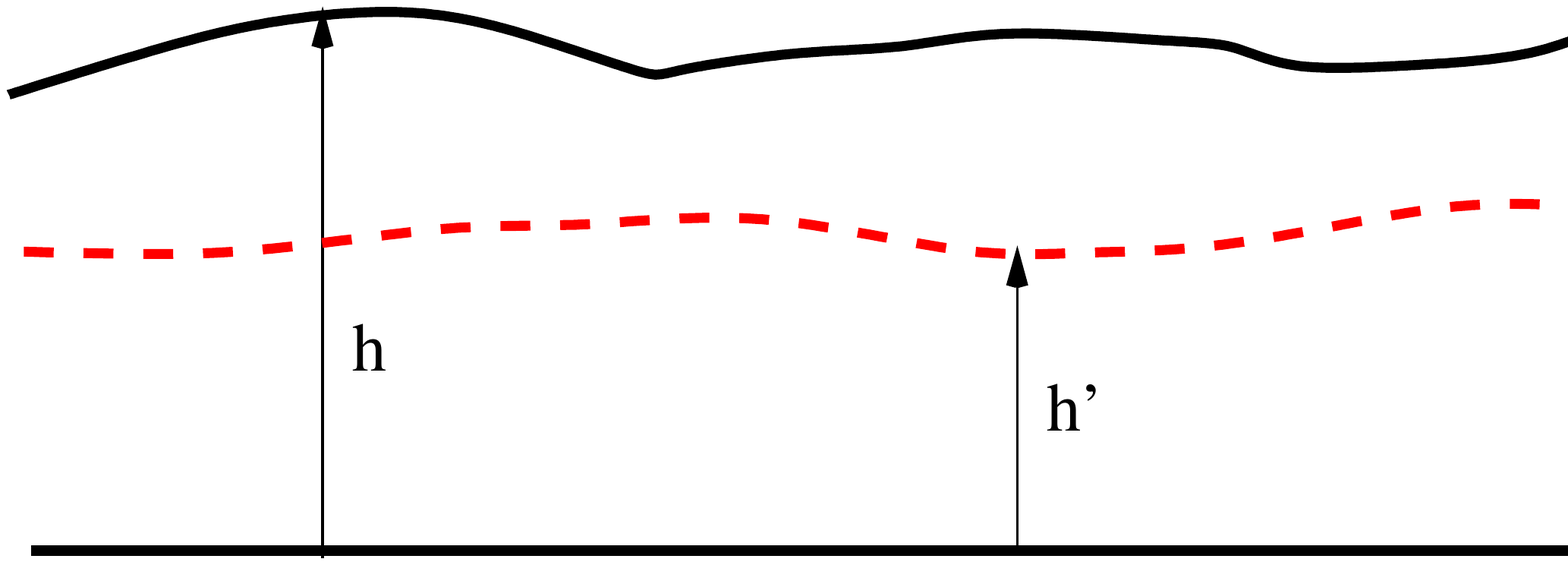}
 \caption{Schematic side view of the lipid membrane (black continuous line) of 
height $h({\bf r})$ and the spectrin network (broken red line) of height 
$h^\prime ({\bf r}')$, measured with respect to a Monge gauge base plain 
(black straight line); $h({\bf r})-h^\prime ({\bf r})=l({\bf r})\neq const.$ 
(see text).}\label{model4}
\end{figure}

In the main text, we have considered a fixed distance $l$ between the lipid 
membrane and the spectrin network. In a realistic situation, the distance $l$ 
should be a fluctuating quantity; see Fig.~\ref{model4}. In live RBCs, the 
average spacing $\langle l\rangle\sim 30 nm$ and is of the same order of 
magnitude as the root mean squared fluctuation amplitude of the lipid 
membrane~\cite{distance}. This should make the fluctuation of $l$ significantly 
affect the fluctuation of the lipid membrane in a live RBC. For an 
ATP-depleted RBC or an artificial system of spectrins deposited on a lipid 
bilayer, the  root mean squared fluctuation amplitude of the lipid membrane 
should be small ($\ll \langle l\rangle$), and hence, should be irrelevant. We 
demonstrate this in a simple model calculation below.

We start with a simple model free energy
\begin{eqnarray}
 {\mathcal F}_{hh^\prime}=\int d^2r [\frac{\kappa_0}{2}(\nabla^2 h)^2 
+\frac{\lambda}{2} u_{ii}^2 +\mu u_{ij}^2 + \tilde A g(h-h^\prime).
 \label{fhhprime1}
\end{eqnarray}
Here, $g(h-h^\prime)$ models the inter-membrane (i.e., between the lipid 
bilayer and the spectrin network) potential; this is a {\em nonuniversal} 
function, and depends on the detailed interactions of the specific lipid 
molecules and spectrin filaments. Nonetheless, on simple physical grounds, we 
impose $g(h-h^\prime)=0$ for $h=h'$ (no interaction potential when the two 
superpose), and $g(h-h^\prime)=0$ again for $h-h^\prime \rightarrow \infty$. We 
expect $g(h-h^\prime)$ to have one minimum at an intermediate distance that is 
the {\em preferred distance $l$} between the lipid bilayer and the spectrin 
network; for live RBCs, $l\sim 30 nm$~\cite{distance}. Parameter $\tilde A$ is a 
function of any quantity that is tilt-invariant; this models the fact that the 
the magnitude of the intermembrane potential not only depends on the local 
distance between the two, but also on the local configurations the lipid 
membrane and spectrin bilayer, which are modeled by $\tilde A$. In the limit 
when the lipid bilayer is completely decoupled from the spectrin network, 
$g(h-h^\prime)=0$ identically.

At a finite temperature $T$, there should be fluctuations in the local distance 
about the preferred distance; the extent of fluctuations should depend on the 
depth and sharpness of the potential well (potential minimum). In a long 
wavelength approximation, choose
\begin{equation} 
 \tilde A=\alpha_1 u_{ii}\nabla^2 h+\alpha_2 u_{ii}\nabla^2 h^\prime,
\end{equation}
where we have ignored possible dependences of $\tilde A$ on 
$u_{ij}^2$,or its linear dependences on $\nabla^2h,\nabla^2h^\prime$ (and any of
their products), as these do not 
affect the line of arguments outlined below; $\alpha_1,\alpha_2$ are 
phenomenological constants. In the fixed distance limit, $h-h^\prime=l_0$ (a 
const.), $g(h-h^\prime) = a_1 l_0$ for small $l_0$, $a_1$ is another 
phenomenological constant. With this, ${\mathcal F}_{hh^\prime}$ in 
(\ref{fhhprime1}) immediately yields the free energy $\mathcal F$ in the main 
text [Eq.~(1) in the main text] with the identification $\chi=(\alpha_1 
+\alpha_2) a_1 l_0$.

We now generalize by allowing small fluctuations in the distance $l$ about 
$l_0$. Let $h({\bf r})-h^\prime ({\bf r})=l=s({\bf r})+l_0\neq 
const.$, 
$\langle s({\bf r})\rangle =0$, is a 
fluctuating quantity. Small fluctuations of $l({\bf r})$ implies $\sqrt{ 
\langle s({\bf r})^2\rangle} \ll l_0$. For small fluctuations, we write
\begin{equation}
 g(h-h^\prime)=g(l_0+s({\bf r}))\approx g(l_0) + \gamma s({\bf r}),
\end{equation}
where $\gamma$ is a phenomenological constant that we set to unity below 
without any loss of generality.
In that 
case, in terms of 
$h^\prime$, the local strain tensor $u_{ij}$ is given by
\begin{eqnarray}
 u_{ij}&=&\frac{1}{2}[\partial_i u_j + \partial_j 
u_i+(\partial_ih^\prime) (\partial_jh^\prime)]\nonumber \\ &=&
\frac{1}{2}[\partial_i u_j + \partial_j 
u_i+(\partial_ih) (\partial_jh) - \partial_i h\partial_j s-\partial_i 
s\partial_j h +\partial_i s\partial_j s].
\end{eqnarray}
Further assume $\langle s({\bf r})s(0)\rangle=D\delta({\bf r})$. This is a 
reasonable 
assumption, given that there are no long-range microscopic fluctuating degrees 
of freedom expected to be present in thermal equilibrium. This implies 
$C_s=\langle |s({\bf q}|^2\rangle = (2\pi)^d D$, a finite constant even in the 
infra-red limit (wavevector ${\bf q}\rightarrow 0$). Compare this with the bare 
correlator of $h$, $C_h=\langle |h({\bf q}|^2\rangle =1/(\kappa q^4)$, that 
diverges in the limit ${\bf q}\rightarrow 0$. Averaging over $s$ then yields 
additional diagrams which correct $\kappa,\,A$ and $B$. These diagrams have 
exactly the same form as the corresponding one-loop diagrams in the fixed distance 
limit; see Fig.~\ref{kappa_dia-all}, Fig.~\ref{Adiag} and Fig.~\ref{Bdiag}, 
respectively, except that one or more internal lines in the new diagrams now 
correspond to $C_s$. Since $C_s$ is a constant, whereas $C_h$ is infra-red 
divergent, all the new diagrams are {\em subleading} (in a scaling/RG sense) to 
the corresponding diagrams Fig.~\ref{kappa_dia-all}, Fig.~\ref{Adiag} and 
Fig.~\ref{Bdiag}, respectively. Thus, no new relevant corrections are generated.

If we consider the instability due to asymmetry at a finite scale $\xi$, then 
the 
additional one-loop contributions to $\kappa,\,A,\,B$ that are generated after 
averaging over $s$ can shift the threshold on $\chi^2$ for the 
instability (or, the vanishing of $\kappa_e$). However, the qualitative picture 
remains unchanged. We, therefore, conclude that small fluctuations in the 
membrane-elastic 
network does not affect our results in the main text, obtained with the 
assumption of a fixed distance, in any significant manner.


\begin{thebibliography}{99}
 \bibitem{weinberg} {\em Statistical Mechanics of Membranes and Surfaces}, edited
by D. Nelson, T. Piran, and S. Weinberg World Scientific,
Singapore (1989).
 \bibitem{Pontes} Pontes B, Ayala Y, Fonseca ACC, Romão LF, Amaral RF, 
 Salgado LT, et al. (2013) Membrane Elastic Properties and Cell Function. 
 PLoS ONE 8(7): e67708. doi:10.1371/journal.pone.0067708
 \bibitem{Giess} F. Giess {\em et. al}, {\em Biophys. J.}, {\bf 87},  (2004).
 \bibitem{munro} J. R. C. Whyte and S. Munro, {\em Journal of Cell Science},
 {\bf 115}, 2627 (2002).
 \bibitem{wiese} K. J. Wiese {\em Eur. Phys. J. B}, {\bf 1}, 269 (1998).
 \bibitem{md-tethered} F. F. Abraham, W. E. Rudge and M. Plischke, {\em Phys. Rev. Lett.}
 {\bf 62} 15  (1989).
 \bibitem{mcs} Y. Kantor {\em et al}, {\em Phys. Rev. A} {\bf 35} 3056 (1987).
 \bibitem{moldovan} D. Moldovan and L. Golubovic {\em Mat. Res. Soc. Symp. Proc.}, {\bf 543} (1999).
 \bibitem{sinner} Eva-K Sinner and W. Knoll, {\em Current Opinion in Chemical Biology},
 {\bf 5}, 705 (2001).
 \bibitem{joanny1} T. Betz {\em et al}, {\em Proc. Natl Acad. Sci. USA} 
{\bf 106}, 15320 (2009).
 \bibitem{joanny2} H. Turlier {\em et al}, {em Nature Phys.} {\bf 12}, 513 
(2016).
 
 \bibitem{chaikin}  P. M. Chaikin and T. C. Lubensky, {\em Principles of
condensed matter physics}, (Cambridge University Press, Cambridge
2000). 
 \bibitem{peliti2} E. Guitter {\em et al}, {\em J. Phys. 
France} {\bf 50}, 1787 
(1989).
\bibitem{luca} L. Peliti and S. Leibler, {\em Phys. Rev. Lett.} {\bf 54}, 1690 
(1985).
 \bibitem{artificial} I. Lopez-Montero, R. Rodriguez-Garcia and F. Monroy, {\em 
J. Phys. Chem. Lett.} {\bf 3}, 1583 (2012).
\bibitem{safran} T. Auth, S. A. Safran and N. S. Gov, {\em Phys. Rev. E} {\bf 
76}, 051910 (2007).
 \bibitem{alberts} B. Alberts, D. Bray, J. Lewis, M. Raff, K. Roberts, J.D.
Watson, Molecular Biology of the Cell, 3rd edition (Garland,
New York, 1994).
\bibitem{joanny} { We ignore any random attachment-detachment of the 
spectrin layer with the lipid membrane as those are believed to be of 
nonequilibrium origin; see, e.g. Refs.~\cite{joanny1,joanny2} above.}
 \bibitem{operator} We do not distinguish between $\delta_{ij}$ and 
$\delta^i_j$. Their
 differences in the Monge gauge contribute to corrections higher order in 
${\boldsymbol\nabla}h$
 and are ignored here in the long wavelength limit.
\bibitem{nelson-peliti} D. R. Nelson and L. Peliti, {\em J. Physique} {\bf 48}, 
1085 (1987).
 
 
 
 
 
 \bibitem{mismatch} This is quantitatively different from the RG results on 
symmetric tethered membranes, see, e.g., Ref.~\cite{chaikin}. We believe 
that
this quantitative difference is due to different $d$-dimensional generalization of the 
$2d$ theory.
 \bibitem{surface-tension} A clear signature of surface tension in an RBC 
remains controversial till date; see, e.g., W. Choi, J. Yi and Y. W. Kim, {\em 
Phys. Rev. E} {\bf 92}, 012717 (2015).

\bibitem{tb-future} T. Banerjee and A. Basu, work in progress.


 \bibitem{toole} P. J. O'Toole, I. E. G. Morrison and R. J. Cherry,
 {\em Biochimica et Biophysica Acta - Biomembranes}, Elsevier {\bf 1466} 39 (2000).
 \bibitem{grzybek} M. Grzybek {\em et. al}, {\em Chemistry and Physics of Lipids},
 {\bf 141}, 133 (2006).
 \bibitem{hendrich} A. B. Hendrich, K. Michalak, M. Bobrowska and A. Kozubek,
 {\em Gen. Physiol. Biophys.}, {\bf 10}, 333 (1991)
 
 \bibitem{abhijit} M. Mitra, A. Patra and A. Chakrabarti, {\em Eur. Biophys. 
J.} 
{\bf 44}, 635 (2015).
 
 \bibitem{peng} Z. Peng {\em et. al}, {\em Proc Natl Acad Sci U S A}, {\bf 110}, 
33 (2013).
\bibitem{nelson-new} A. Kosmrlj and D. Nelson, arXiv:1606.06750.
\end{thebibliography}

\begin{thebibliography}{99}
 \bibitem{distance} A. Ziker, H. Engelhardt and E. Sackmann, {\em J. Phys. 
(Paris)} {\bf 48}, 2139 (1987); B. S. Bull, R. S. Weinstein and R. A. Korpman, 
{\em Blood Cells} {\bf 12}, 25 (1986); V. Heinrich {\em et al}, {\em Biophys. 
J.} {\bf 81}, 1452 (2001); G. Popescu {\em et al}, {\em J. Biomed. Opt. Lett.}, 
{\bf 10}, 060503 (2005).
\end{thebibliography}
\end{document}